\PassOptionsToPackage{table,xcdraw}{xcolor}
\documentclass[sigconf]{acmart}

\usepackage{multirow}

\AtBeginDocument{%
  \providecommand\BibTeX{{%
    Bib\TeX}}}

\copyrightyear{2023}
\acmYear{2023}
\setcopyright{acmlicensed}\acmConference[FDG 2023]{Foundations of Digital Games 2023}{April 12--14, 2023}{Lisbon, Portugal}
\acmBooktitle{Foundations of Digital Games 2023 (FDG 2023), April 12--14, 2023, Lisbon, Portugal}
\acmPrice{15.00}
\acmDOI{10.1145/3582437.3582453}
\acmISBN{978-1-4503-9855-8/23/04}

\acmConference[FDG '23]{Foundations of Digital Games}{April 11-14,
  2022}{Lisbon, Portugal}

\begin{document}

\title{The Right Variety: Improving Expressive Range Analysis with Metric Selection Methods}

\author{Oliver Withington}
\email{o.withington@qmul.ac.uk}
\affiliation{%
  \institution{Queen Mary University of London}
  \city{London}
  \country{UK}}

\author{Laurissa Tokarchuk}
\email{laurissa.tokarchuk@qmul.ac.uk}
\affiliation{%
  \institution{Queen Mary University of London}
  \city{London}
  \country{UK}}

\maketitle
\section{Abstract}

Expressive Range Analysis (ERA), an approach for visualising the output of Procedural Content Generation (PCG) systems, is widely used within PCG research to evaluate and compare generators, often to make comparative statements about their relative performance in terms of output diversity and search space exploration. Producing a standard ERA visualisation requires the selection of two metrics which can be calculated for all generated artefacts to be visualised. However, to our knowledge there are no methodologies or heuristics for justifying the selection of a specific metric pair over alternatives. Prior work has typically either made a selection based on established but unjustified norms, designer intuition, or has produced multiple visualisations across all possible pairs. This work aims to contribute to this area by identifying valuable characteristics of metric pairings, and by demonstrating that pairings that have these characteristics have an increased probability of producing an informative ERA projection of the underlying generator. We introduce and investigate three quantifiable selection criteria for assessing metric pairs, and demonstrate how these criteria can be operationalized to rank those available. Though this is an early exploration of the concept of quantifying the utility of ERA metric pairs, we argue that the approach explored in this paper can make ERA more useful and usable for both researchers and game designers.

\section{Introduction}

Procedural Content Generation (PCG) systems for games can often produce huge amounts of diverse artefacts and experiences. Whether it is the billions of possible maps of Minecraft \cite{persson2011} or the quintillions of planets in No Man Sky \cite{hellogamesfirm2016}, the total volume of a PCG system’s possible output, commonly referred to as its ‘generative space’ can be huge and complex. The challenge of understanding and usefully comparing generative spaces is a vital one to both games researchers and game designers. Researchers want to be able to credibly claim that their novel systems are an improvement over prior alternatives in terms of the diversity or quality of output. Similarly, commercial game designers want to be confident that they are selecting the right generative approach, and tuning their chosen approaches correctly to produce the optimal generative space. 

A popular approach for gaining an understanding of a generative space, is to produce visualisations of it which condense the highly convoluted n-dimensional spaces of possible artefacts, into something 2-dimensional and human understandable.
A common technique for producing 2D visualisations of generative spaces in prior PCG research is Expressive Range Analysis (ERA), a technique pioneered by Smith et al \cite{smith2010}. ERA allows us to visualise sets of game levels using two or more quantifiable metrics of interest such as heuristics for difficulty or aesthetic qualities. These metrics are used to localise each artefact in a 2-dimensional plot, which can then be used to understand how the generated content is distributed in this metric defined space. This helps designers to understand how the generative space of their generator is positioned within the larger possibility space of levels that could be produced within that domain, and to compare that generative space to alternatives.

Arguably the most important choice when applying ERA is the pair of metrics that are used to produce the plot. However, when ERA has been deployed in prior research this choice is typically not justified, and the impact of it is normally not discussed. There are to our knowledge no prior works that aim to justify the value of specific metric pairs over the alternatives available, or methods to assist a researcher or designer in deciding on the best pair to use. Most commonly, research using ERA either repurposes metric pairings from prior work, often Smith et als original paper, or a large number of ERA plots is produced for all metric pairs. 

It is to this area of selecting metrics for applying ERA that this paper makes a contribution. We argue that there are actually several characteristics of metric pairs which could make them more desirable than alternatives when applying ERA, and that the calculation of these characteristics across candidate pairs of metrics can form the basis of helpful selection criteria. Specifically the selection criteria we look at are: 

\begin{itemize}
\item Fitness Independence, meaning that fit content is evenly distributed in the metric pair defined space
\item Mutual Independence, meaning there is minimal correlation between the pair of metrics selected
\item Minimal Alternative Metric Independence, meaning there is minimal metric diversity which is unaccounted for by a selected metric pair
\end{itemize}

This work makes the case that the more a metric pair satisfies these criteria, the more useful an ERA visualisation using them will be. Combined, they help to ensure that as much of the ERA plot as possible is accessible for the generator, that as much of the plot is useful for the generator to explore, and that what appears to be diversity for generative spaces using it actually represents diversity across many possible metrics. 

In this paper we introduce these selection criteria for ERA metric pairs and discuss their value to both game designers and researchers looking to understand or compare their systems’ generative spaces. We show how these characteristics can be calculated to differentiate between and rank alternative metric pairings, allowing a more informed choice to be made about which pair to visualise in cases where it is not obvious. We also conduct a series of pilot experiments in the popular research domain of generated levels for Super Mario Bros, using the approach to evaluate and rank over 150 possible metric pairings for visualising generated output using ERA. The results of these experiments suggest that in spite of being an early exploration of these ideas, applying these selection criteria to metric selection can be a valuable source of insight into how best to conduct ERA within specific game domains, as well as potentially a source of more general insight into the design of good metrics. We conclude that this approach could substantially improve on the usability and utility ERA as it is currently applied while maintaining its benefits, allowing it to be used as a more robust approach for understanding and comparing the generative spaces of PCG systems.

\section{Related Work}\label{sect_relwork}

In this section we introduce and discuss the work that most directly informed and inspired this project, both to highlight the core concepts that underpin it and to show where this project aims to improve on the current state-of-the-art.

\subsection{Expressive Range Analysis}

ERA was first introduced by Gillian Smith and Jim Whitehead in 2010 \cite{smith2010}, and was proposed as a novel method for visualising and understanding generative spaces. Its mode of operation is appealingly simple, which has no doubt contributed to its widespread adoption. To conduct basic ERA a designer first generates a large set of levels from a generator, and then calculates two or more metrics for each level. The set of levels can then be visualised on a 2D plot, most commonly a heatmap. These 2D plots can then be used to understand the character and diversity present in individual generative spaces, but also those from alternative generators compare to each other. ERA has most commonly been applied to work within its original domain of PCG for game levels, but it has also found a place in alternative areas of PCG research such as visualising the output of poetry generators \cite{kreminski2022}, tabletop game outcomes \cite{guzdial2020} and emergent narrative systems \cite{kybartas2018}. 

Over the intervening years ERA has been used for several alternative purposes within PCG research. In Smith et al’s original paper \cite{smith2010} it was primarily used to qualitatively understand their systems output tendency, with a focus on characteristics such as what kinds of levels are likely to be produced and which levels are hard or impossible to produce. However they also highlighted that ERA could form the basis of quantitative comparisons between generators, a use case that has proved very popular in research since. It has often been used to heuristically assess the relative amounts of output diversity which can be produced from alternative systems. Output diversity is often regarded as an inherently desirable quality of PCG systems, and ERA provides a way of comparing the output diversity of alternative systems by comparing which generative space is more spread out in the ERA plot. As a result it commonly appears in novel PCG research to support arguments about how PCG systems compare in terms of the diversity present in their generative spaces \cite{horn2014,jadhav2021a,zakaria2022,alvarez2022}.

Though ERA is still used in contemporary research in its original form, there have been several evolutions on the underlying concept to improve its applicability and functionality. Adam Summerville introduced the concept of using corner plots to visualise multiple metric pairs simultaneously, overcoming the prior limitation of only viewing two metrics at a time \cite{summerville2018}. They also introduced the use of the e-distance metric for comparing the distributions of generative spaces in n-dimensional space. Cook et al developed the Danesh system for conducting automatic ERA within Unity, as well as for identifying the trends in ERA across alternative parameterisations of the same generator using what they termed Randomised ERA, or RERA \cite{cook2021}. Kremenski et al introduced ‘Expressive Range Coverage Analysis’ to visualise trends in how designers interacted with a generative system over time \cite{kreminski2022}. We have also explored the use of alternative techniques such as dimensionality reduction and Convolutional Neural Network embeddings to produce ERA style generative space visualisations without the need for defining or recalculating the type of diversity we are interested in\cite{withington2022a,withington2022b}.

\subsection{Generative Space, Possibility Spaces and ERA}\label{sect_pspace}

An important concept for this work, but also ERA more broadly, is that of generative spaces and possibility spaces. These terms were popularised by Mike Cook in their work \cite{cook2022} and they help to highlight the strengths and weaknesses of ERA, as well as the potential benefits of this work’s approach. Unfortunately the definition of these terms is not universally accepted, and there are high profile works that use generative space and possibility space interchangeably\cite{guzdial2020}. The distinction remains very conceptually useful though, and we hope the terminology will become more ubiquitous in future.

As we have previously discussed, the generative space of a PCG system refers to the volume that contains every artefact that a specific system could generate. In contrast, possibility space refers to the volume that contains every artefact that could be represented by any generator for that domain. Typically, a generative space is a smaller volume contained within the larger possibility space. When a game designer is choosing between alternative generative approaches for a certain task, or tuning the parameters of a chosen generative approach, this can be conceptualised as the choosing of an optimal generative space for a given possibility space. Similarly, when a designer is comparing two generators in terms of their output, they are comparing two different localisations of generative spaces within the same possibility space.

This terminology gives us a lens with which to understand ERA. ERA plots simultaneously produce a 2D dimensional projection of both a PCG systems generative space, and also the domains underlying possibility space. Just as the n dimensional volume of the generative space is compressed into a 2D metric determined plot, so is the potentially vastly more complex underlying possibility space. The utility of an ERA visualisation of a generative space is heavily influenced by the impact of the selected metric pair on the projection of the possibility space. When conducting ERA it is very possible to decide on metric pairs which mean that the majority of the plot does not contain good content, or is not even within the possibility space of content due to the majority of metric values being mutually incompatible. The intention with this work’s approach is for it to help designers use ERA to be a more useful projection of the underlying possibility space, and therefore a more useful projection of a specific generative space, while maintaining the explainability and simplicity of ERA.

\subsection{Metric Selection for ERA}

The main decision a designer has to make when applying conventional ERA to a generative space is the choice of the pair of metrics to visualise the space. This choice dictates the types of diversity that will be captured in the visualisation, as well as potentially what types of diversity are ignored. The distribution of any metric that is of interest but is not included in the plot is unknown. Given three metrics of interest, a plot could show a high level of diversity in terms of metrics 1 and 2, while concealing homogeneity in metric 3 for example. If ERA is being deployed to allow a designer to qualitatively explore the distribution of a generative space in terms of two metrics that they know are of interest, then no choice needs to be made. However, if ERA is being deployed to argue that a generator produces more generally diverse content than an alternative, or if there are many metrics of interest for a designer, then this choice is less obvious.

In general there are two approaches that are taken for selecting metrics. Firstly, researchers can reuse metrics that have been used in prior literature, most commonly the two metrics from Smith et als original paper. In that paper Smith et al introduced a ‘linearity’ metric, which is a heuristic for how flat a level is, and ‘leniency’, a heuristic for level difficulty. Though these two metrics were designed to be insightful within the context of their specific level generation system, they have since been widely reused within PCG research for conducting ERA  \cite{horn2014,alvarez2022,jadhav2021a}, and also as metrics to intentionally aim to find diversity in as part of Quality-Diversity search based PCG \cite{alvarez2019,sarkar2021}. While there is nothing inherently problematic about reusing metrics in this way, the lack of justification given for their reuse in novel domains means researchers risk using suboptimal metrics for their purposes.

We also note that while the names linearity and leniency appear frequently to describe metric choices, the underlying calculation to arrive at them is often altered to fit with the specific game domain of interest. Leniency was first calculated as a score based on the number of jumps, moving platforms and enemies in the level \cite{smith2010}, but has since been calculated as a normalised rating on the average difficulty of every point of the level \cite{horn2014}, or more simply as count of enemies subtracted from count of rewards \cite{jadhav2021a}. Linearity similarly was originally calculated as a linear regression on the gaps between the highest points of the level for each horizontal position and the centre line \cite{smith2010,shaker2012}, but has since been calculated in very different ways in different contexts, such as using a normalised count of the number of paths between doors in the case of Alvarez et al’s work on the Evolutionary Dungeon Designer (EDD) \cite{alvarez2022}. These alternative calculations usually make pragmatic sense. The EDD for example creates levels that are viewed top down and contain no deadly gaps in the scenery which makes the original formula inappropriate. However, the degree of variation in even these ostensibly similar metric pairs for ERA highlights, in our view, the potential value of heuristics for making a more informed choice.

The alternative approach for metric selection is to not select at all, and instead visualise the plots for all metric pairs that might be of interest. This longer list of metrics can again be sourced inclusively from the prior literature on similar game domains \cite{alvarez2022,herve2021}, or a list can be designed by a game domain expert \cite{kreminski2022}. This list of n metrics can then be visualised using ERA of all possible pairs, commonly using Summerville’s corner plot concept \cite{summerville2018}. However, while presenting all possible visualisations limits the impact of one of them being misleading or unideal, it does not avoid them being present. It also can arguably limit the readability of presenting a single plot, though this is not necessarily the case.

There have been efforts to add more robust justification for choosing individual metrics within ERA, most commonly in the form of validating whether or not commonly used metrics correlate with human perceptions of quality and diversity. Summerville et al explored the power of common metrics when predicting player experience and found strong correlations \cite{summerville2017}. Marino et al conducted a similar study and found much more limited correlation \cite{marino2021}. Herve \& Salge explored the relationship between common metrics and expert evaluations of Minecraft maps \cite{herve2021}. These works have been invaluable for confirming or refuting that specific metrics are capturing forms of quality or diversity which are of real interest to designers. However, as their focus has been on the informative power of individual metrics their utility in aiding the selection of metric pairings is limited. As we shall see in Section \ref{sect_approach}, applying ERA with two metrics which are strongly correlated with human perceptions does not ensure the plot will be useful. It is still possible for two such metrics to produce an unhelpful projection of the possibility space, one in which for example, the majority of the space is undesirable or inaccessible.

\subsection{Mario AI Benchmark}

This work’s pilot experiments make use of the Mario AI Benchmark, a widely used AI research platform used for the development and evaluation of AI agents for level playing and level generation. While we share the sentiment of Mike Cook that Super Mario has received an unreasonable amount of focus within PCG literature as a test bed \cite{cook2022} , using it within this work is justified by the amount of prior work that has applied ERA specifically to Mario level generators. This gives us both a wide variety of previously used metrics to draw upon, as well as making any insights we gain into optimal metric selection more relevant to the research community.

We use the level corpuses generated as part of Horn et al’s 2014 work \cite{horn2014}, generated from a diverse set of nine Mario level generators which are included in the latest version of the benchmark, which is bundled with 15 levels from the original Mario game giving a total of 9015 levels. These level sets are ideal for this research for several reasons. They all use the same encoding system in which levels are represented by 2D matrices of characters, in which each character represents a tile type. This means we can use the same metric calculation approaches across all levels in the corpus. Additionally, prior research has confirmed that the sets from the alternative generators are meaningfully diverse in a way that has been detectable using alternative visualisation approaches\cite{horn2014,withington2022a}.

To automatically evaluate the levels we make use of Robin Baumgarten’s A*-based Mario level playing agent, the winning agent in the first Mario AI competition \cite{togelius2010}, which is included in the benchmark and is often used in PCG research to confirm that a level is completable \cite{fontaine2021b,cernygreen2020}. 

\section{Metric Pair Selection Criteria}\label{sect_approach}

As we have seen there is little consistency in how ERA has been applied in terms of the metrics selected, and that there is typically limited justification given for the selection of certain metrics and pairings over alternatives. However, we argue that there are characteristics of useful ERA plots that we can infer from its design, characteristics that can be used as the basis for heuristics for guiding metric pair selection. In this section we introduce and discuss the relevance of these characteristics, as well as how they can be operationalised to produce metric pair selection criteria. All of the selection criteria we introduce are intended to be calculated for a given pair of metrics, selected from a larger pool of candidate metrics. For a set of n metrics that may be of interest, there are n(n-1)/2 potential pairs of metrics which could be selected to produce ERA plots. The goal is for these selection criteria to give a quantifiable justification for selecting one of these potential pairs over the alternatives. 
\subsection{Criteria 1: Fitness Independence}
\begin{figure}
  \centering
\includegraphics[width=\linewidth]{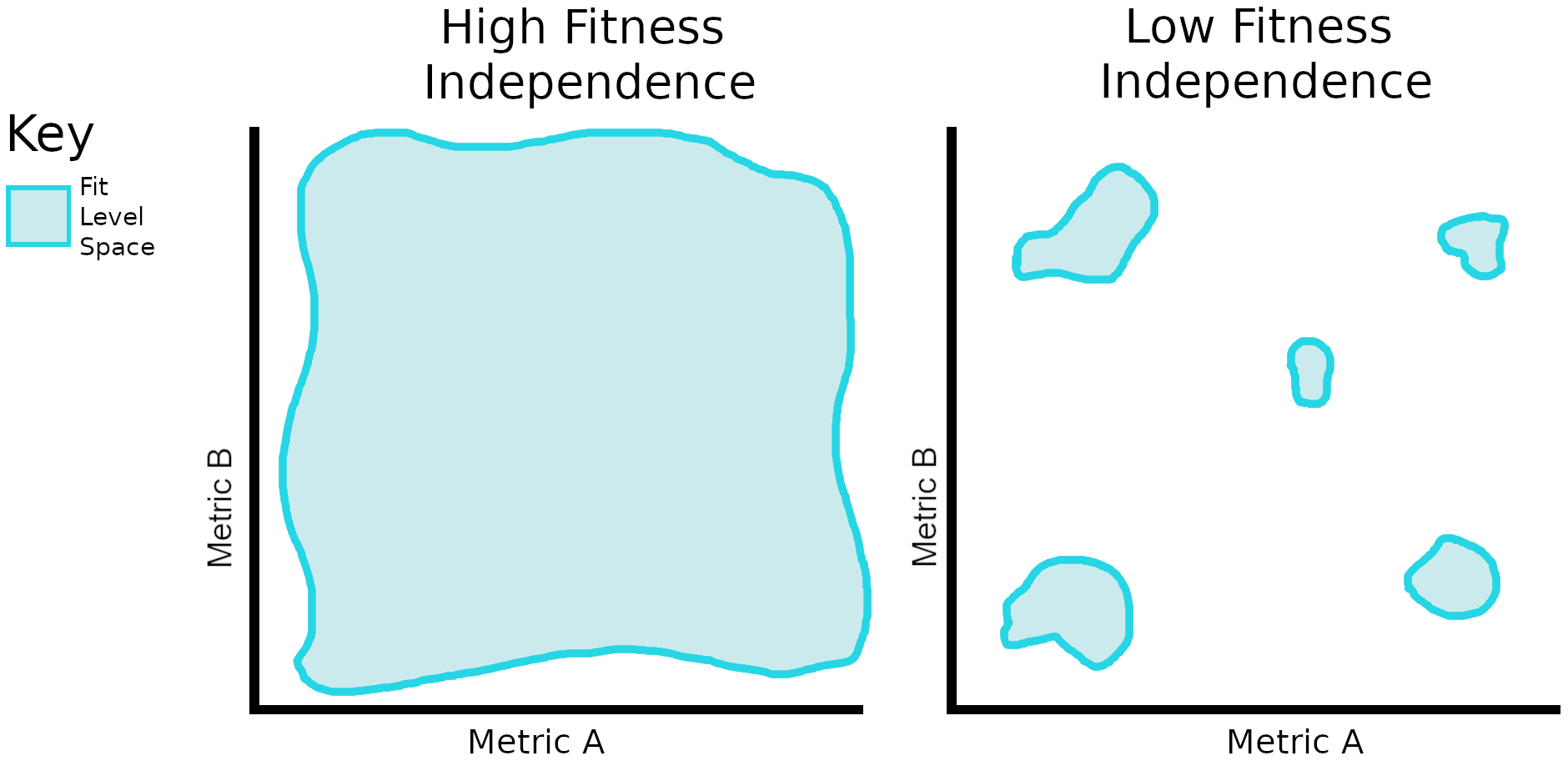}
  \caption{Illustration of the presence of high and low fitness independence in an ERA space in terms of a single alternative metric}
  \label{fig_fitind}
\end{figure}

The first selection criteria for metric pairs that we argue to be valuable is that of fitness independence. What we mean by fitness independence is that fit or desirable levels are as evenly distributed within the 2D ERA projection of the underlying possibility space as possible (See Figure \ref{fig_fitind} for an illustration). Fitness is a term popularised by the field of search-based algorithms to refer to quantifiable measures of how good or strong an artefact is. It is important to note that this criteria is naturally only applicable to domains where there is a concept of fitness, though in the domain of generating game levels such a concept is extremely common due to how easy it is for PCG systems to generate game levels that are uncompletable, or only completable to a point. Fitness in PCG for game levels is often contextualised as a more binary concept  \cite{khalifa2018a,gallotta2022}, and we consider that this fitness independence selection criteria would be useful both for domains with either continuous or binary fitness concepts. Either way, for exploration within ERA to be desirable, there needs to be a good distribution of acceptable content.

We argue that this criteria is implicitly desirable for an ERA visualisation to be as informative as possible. With it, alternative generative spaces can be directly compared in terms of their relative explorations of the space, or an individual generator can be tuned to locate its generative space wherever the designer wishes within the space, while being confident that each location in the space is equally likely to contain fit content. Contrastingly, in visualisations with low fitness independence exploration is likely to be undesirable, and a designer would not feel as free to optimise towards any chosen location of the space, as all of the fit levels would be tightly clustered in the space. An ERA plot with low fitness independence could still be useful to a designer confident that the two metrics represent the diversity they are most interested in, but they would still need to be more wary of the location of fit content within the space. High fitness independence reduces the need to be concerned about the location of fit content, allowing researchers and designers to focus more just on the visualisation.

The method for calculating the fitness independence that we explore in this paper is to discretise the ERA plot into an evenly sized grid, in which each cell in the grid is defined by a range of metric values. Each level to be visualised is then sorted into this grid. We can then calculate the average fitness of each grid cell in terms of the levels placed there, and then further calculate the average cell fitness for the entire grid. Low values for a grid defined by a certain metric pair would therefore indicate that fit levels were highly concentrated in specific cells of the grid with more of the grid dominated by either empty cells or cells with exclusively low fitness scores, whereas higher values would indicate a more even distribution.

\subsection{Criteria 2: Mutual Correlation}

\begin{figure}
  \centering
\includegraphics[width=\linewidth]{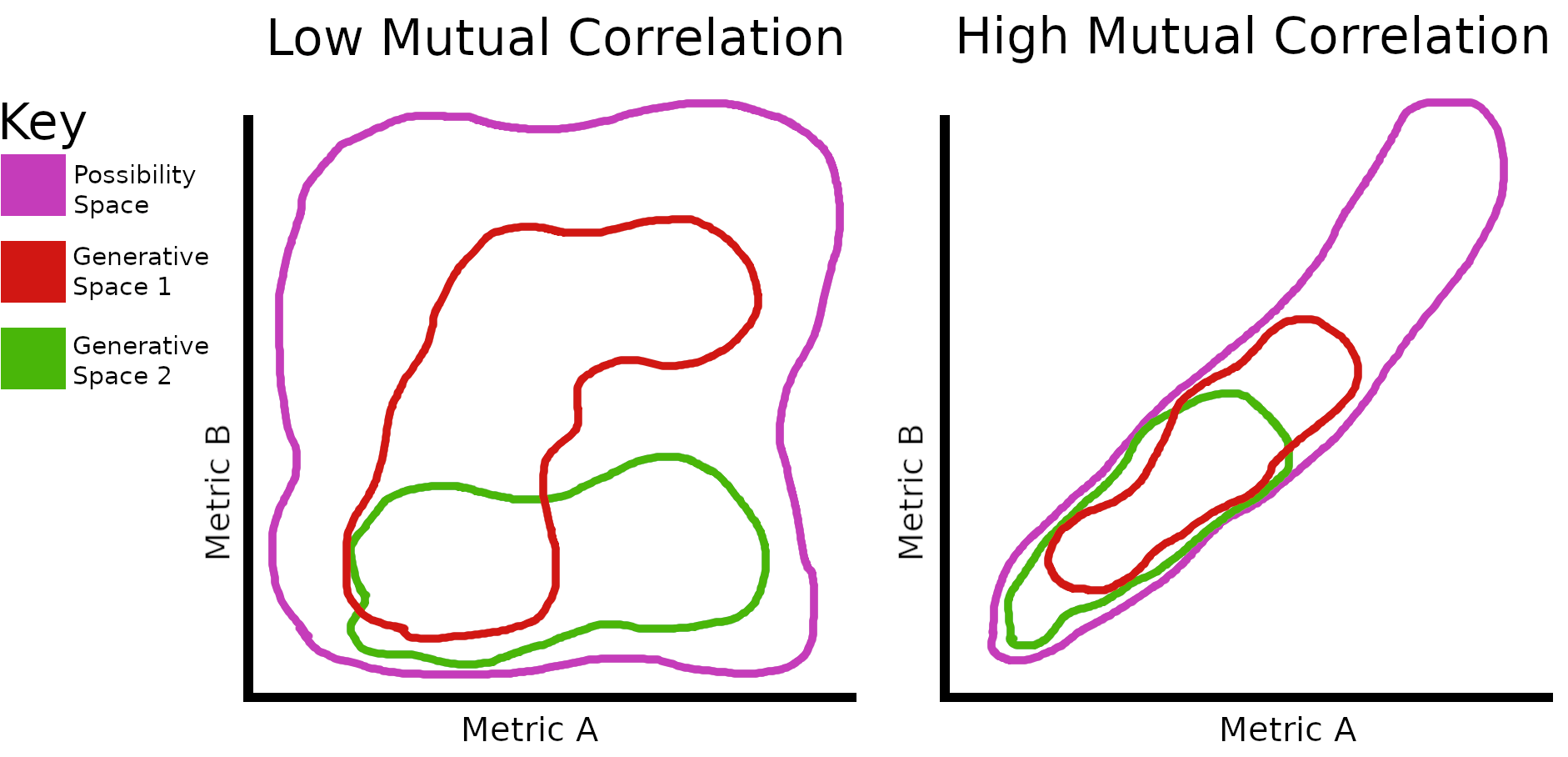}
  \caption{Illustration of the effect of low and high mutual correlation on the location of a possibility space within an ERA plot, as well as the effect on the location of generative spaces}
  \label{fig_metcorr}
\end{figure}

The second selection criteria for metric pairs is that of mutual correlation. What this means practically is that we want there to be a minimal or non-existent predictive relationship between the values of one metric and the other in its pair (See Figure \ref{fig_metcorr} for an illustration of this criteria). This criteria is important for two reasons. Firstly, to ensure that as much of the plot is actually accessible, in other words that as much of the plot is within the underlying possibility space of the domain. Secondly, it is important to minimise the amount of redundant information within the ERA plot. In the case that the value of one metric is highly predictive of the other, then one could be discarded while maintaining the same amount of information about the distribution of the generator’s levels. In the alternative desirable case in which there is minimal correlation between the metric values in a selected pair we are instead maximising the information gained within the plot, and also ensuring that more of the plot is theoretically accessible. 

Meeting this criteria makes the ERA plot using the pair more meaningful. With the criteria met we can be confident that what appears as low diversity in an ERA plot means that a generator is exploring a small amount of a possibility space, not that the projection of the possibility space itself is heavily restricted. It also makes it more likely that a designer will be able to distinguish between the properties of alternative generative spaces. The stronger the mutual correlation between metrics, the more that any projected generative spaces would necessarily overlap in the amount of the plot that represented the restricted possibility space. As we shall discuss in the next section on the third and final criteria, low mutual correlation between a metric pair also increases the likelihood that one of them is correlated with a given alternative metric, a feature we shall introduce as desirable shortly.

To calculate the mutual independence of metrics in our pilot studies we calculate the linear correlation between the values found for each metric using Spearman’s rank correlation coefficient, commonly referred to as Spearman’s $\rho$. The stronger the correlation between the two metrics the more the value of one determines the other, and the less explorable the full ERA plot will be. We first store the absolute value of Spearman’s $\rho$, as whether the correlation is strongly positive or strongly negative we consider it to be equally undesirable. We then subtract it from 1, giving a final score between 0 and 1 except with 1 indicating no correlation, bringing it inline with the other two metrics where low values are less desirable than high. The major disadvantage of using Spearman’s $\rho$ is that it does not help us discover non-linear correlations between metrics which if they were present would still be harmful to how explorable the plot is, something we discuss further in Section \ref{sect_limitations}. However it has the advantage of being far simpler and less computationally intensive to calculate than methods for investigating non-linear correlation.

\subsection{Criteria 3: Alternative Metric Correlation}

\begin{figure}
  \centering
\includegraphics[width=\linewidth]{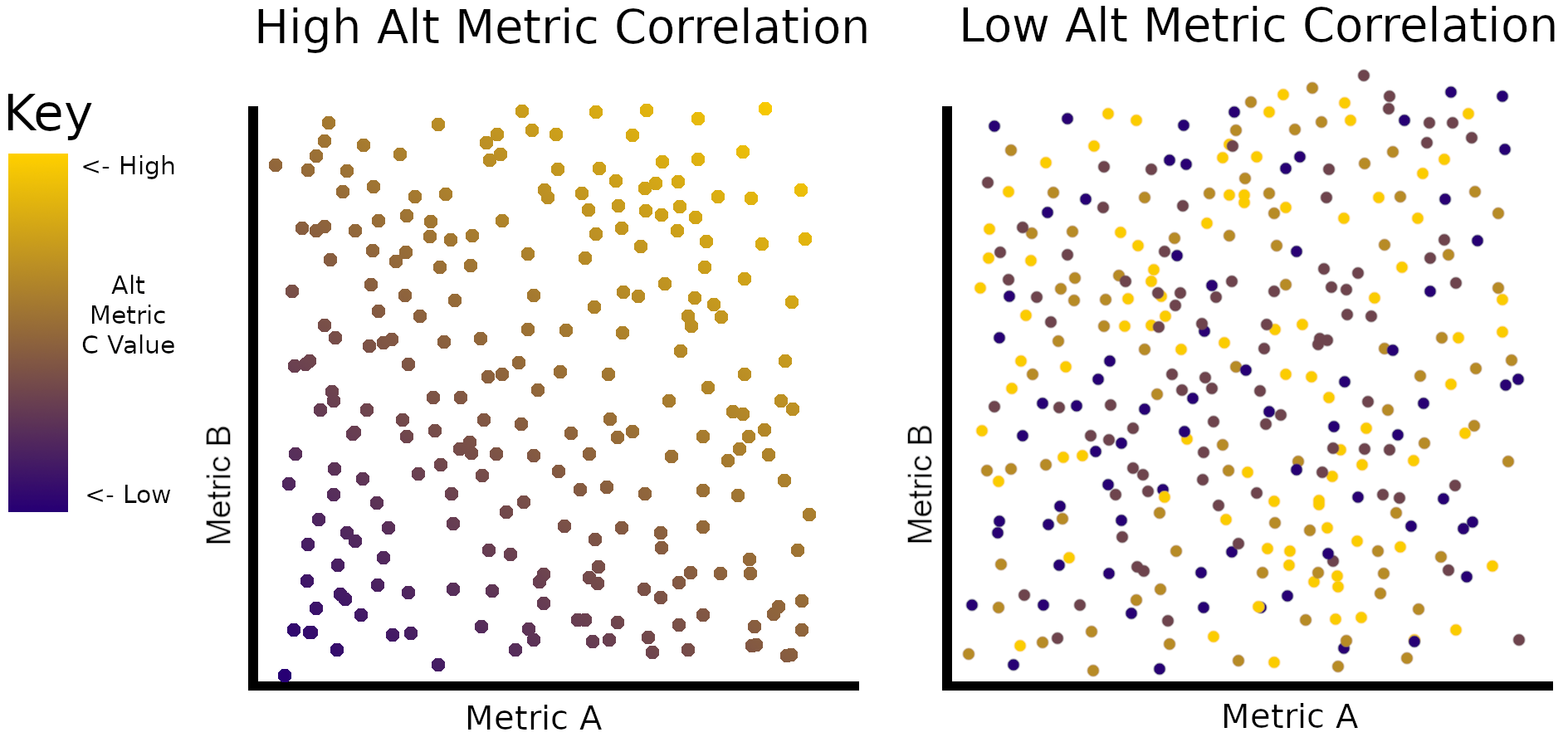}
  \caption{Illustration of the effect of high and low alternative metric correlation in terms of a single alternative metric, and the effect this can have on the distribution of alt metric values in and ERA plot}
  \label{fig_altmet}
\end{figure}

The third selection criteria we investigate in this work is the hardest to conceptualise and calculate, but also arguably the most important. It is that of the level of correlation with alternative metrics. What this means in practice is that we want to minimise the existence of alternative but valuable metrics to the pair used to create the plot, which are uncorrelated with the plotted metrics (See Figure \ref{fig_altmet} for an illustration of high and low alternative metric correlation with a single metric). The underlying goal is that the ERA plot projection of the underlying possibility space is as information rich as possible, and that what appears to be diversity or a lack of it represents diversity or a lack of it in a more holistic way than one determined by just the two metrics.

To understand why this criteria is valuable we shall use a short example case. If we imagine we were interested in only three metrics for a set of generated levels, called A, B and C. If we used A and B as the metric pair to conduct ERA, there is a chance that the generative space could appear widely distributed in the plot even if all content had highly homogenous values for metric C. However, if strong correlation was found between either metrics A or B and metric C, then we could be confident that there was no risk of this, and that what appeared as diversity in A and B also represented diversity in C. This is what we aim to assess with this selection criteria, except that we aim to find minimal independence between the selected pair and all other metrics of interest, rather than with a specific alternative metric. The higher the values for this criteria, the more we can be confident that diversity in an ERA plot represents diversity for multiple metrics, and therefore represents diversity in a more general sense than when the criteria values are low.

To operationalise this criteria we use a similar approach to the mutual independence criteria, except in this case high correlation values for Spearman’s $\rho$ are desirable. For a candidate metric pair M1 and M2 we calculate $\rho$ for the correlation between every alternative metric and M1 and M2 separately, giving two values for each alternative metric. We only store the highest absolute value found for either M1 or M2 as it does not matter which metric in the pair has the correlation, nor whether the correlation is positive or negative. This is why the mutual correlation criteria supports this one, as having two independent metrics increases the likelihood of finding one that performs well for alternative metrics here. This set of $\rho$ values for each other metric can then be divided by the number of alternative metrics to give a mean average correlation score. The higher this score, the more inclusive an ERA plot is of alternative forms of diversity.

\section{Experiment Design}\label{sect_expdesign}
In this section we introduce the design of our pilot experiments using this approach for selecting ERA metric pairs, as well as explaining the reasoning behind the key design decisions. The intention for these experiments is that they demonstrate the potential utility of this work’s novel selection criteria, as well as highlighting the strengths and weaknesses of our current approach for calculating them.

\subsection{Game Domains Assessed}

The PCG generative spaces analysed all come from the Horn et al corpus of generated levels \cite{horn2014} that we discussed in more detail in Section \ref{sect_relwork}. This set of 9000 generated levels from 9 different generators, as well as 15 of the levels from the original Super Mario Bros is included in the current version of the Mario AI Framework. 

When calculating our selection criteria and evaluating which metric pair is most appropriate for performing ERA, we focus on the complete assembled set of 9015 levels as we argue this will give a more representative view of the underlying possibility space of Mario levels than could be obtained from any individual generator in the set. This is a decision we discuss and critique in more detail in Section \ref{sect_limitations}, as while it gives a more nuanced view of the possibility space it is also a less realistic implementation. In practice game designers and researchers will rarely have access to multiple different generators for a game domain to use to gain this kind of view. 

\subsection{Metrics Assessed}

For each of the 9015 Mario levels in our dataset we calculated 18 metrics. For the full list of metrics and how they are calculated see Table \ref{tab_metrics}. The list of 18 metrics and how they were calculated were largely inspired by prior works which used them or similar variants in ERA or QD-search. Metrics related to block counts, or repeated patterns such as locations in which the agent can stand are commonly used, as are counts or ratios of agent actions when simulated play is involved in artefact evaluation \cite{summerville2017,herve2021}. Our MaxJumpAirTime and AverageY metrics are more novel. MaxJumpAirTime captures the largest jump performed by the agent, as performing a large jump can often be a memorable moment or difficulty spike within the level. AverageY relates more to the level’s structure, capturing how the average height of the agent during its run to get an idea of the average height of standable terrain. As this work is focused on metric pair evaluation, the design of the metrics themselves was not a focus. However we did want to have an even amount of ‘structural’ metrics calculated from the level’s structure itself, and ‘agent evaluated’ metrics calculated from the play trace of an AI agent, to draw comparison between the presence of both in high performing metric pairs.

Also extracted from the agent simulations was the amount of horizontal progress the agent was able to make as a proportion of the whole level width. This feature is often referred to as ‘playability’ or ‘fitness’ in the prior work using the Mario AI benchmark\cite{cernygreen2020,fontaine2021b}, though we refer to it as playability as it is the more specific term. Playability is calculated as a score from 0 to 1, where 1 indicates the agent was able to reach the end of the level. This metric was not included as a candidate for inclusion in metric pairs, as it was held back for use as the fitness heuristic in the FI selection criteria calculation.

\begin{table*}
  \caption{Metrics investigated and how they are calculated}
  \label{tab_metrics}
\begin{tabular}{|l|l|l|}
\hline
\textbf{Type}       & \textbf{Name}          & \textbf{How Calculated}                                                                                                         \\ \hline
                    & Contiguity      & +1 per solid tile adjacent to a solid tile                                                                                      \\ \cline{2-3} 
                    & Linearity       & +1 per solid tile horizontally adjacent to a solid tile                                                                         \\ \cline{2-3} 
                    & BlockCount      & +1 per solid tile                                                                                                               \\ \cline{2-3} 
                    & EnemyCount       & +1 per enemy                                                                                                                    \\ \cline{2-3} 
\textbf{Structural} & RewardCount       & +1 per tile containing a reward                                                                                                 \\ \cline{2-3} 
                    & EmptyCount       & +1 per empty tile                                                                                                               \\ \cline{2-3} 
                    & PipeCount         & +1 per tile containing a pipe block                                                                                             \\ \cline{2-3} 
                    & Density          & Total places Mario can stand divided by level length \\ \cline{2-3} 
                    & ClearColumns     & +1 per column containing only empty tiles                                                                                       \\ \hline
                    & Playability            & Proportion of level the agent was able to complete                                                                              \\ \cline{2-3} 
                    & JumpCount        & +1 per jump action used by agent                                                                                                \\ \cline{2-3} 
                    & JumpEntropy       & JumpCount divided by total actions                                                                                              \\ \cline{2-3} 
                    & Speed          & Proportion of level navigated per second on average                                                                             \\ \cline{2-3} 
\textbf{Agent}      & TimeTaken        & Time used by the agent for the run in seconds                                                                                   \\ \cline{2-3} 
\textbf{Extracted}  & TotalEnemyDeaths & +1 per enemy who died during the run                                                                                            \\ \cline{2-3} 
\textbf{}           & KillsByStomp    & +1 per enemy killed by agent by jumping on them                                                                                 \\ \cline{2-3} 
                    & MaxJumpAirTime  & Longest jump by agent in terms of frame count while ungrounded                                                                           \\ \cline{2-3} 
                    & OnGroundRatio   & Proportion of time agent spent on the ground                                                                                    \\ \cline{2-3} 
                    & AverageY         & Average Y axis position of Mario during the run                                                                                 \\ \hline
\end{tabular}
\end{table*}

\subsection{Metric Selection Criteria}

The main results calculated the three selection criteria explained in Section \ref{sect_approach}, namely Fitness Independence (FI), Mutual Correlation (MC) and Alternative Metric Correlation (AMC). Each one gives a score between 0 and 1, with higher scores being more desirable. These three criteria are calculated for each of the 153 unique metric pairs within the set.

Additionally we calculate the rank for each pair in terms of the three criteria, which we can then use to identify the best n metric pairs for each criteria. We also calculate the average rank for each pair across the three criteria. This gives a heuristic for the overall performance of the pair, with lower values indicating better general performance across the three metrics.

\subsection{Agent Simulation}

The agent we use to calculate the agent-evaluated metrics, as well as the playability of the levels is Robin Baumgarten’s A* Mario agent which we introduced in Section \ref{sect_relwork}. This agent has been widely used to assess the quality and characteristics of Mario levels in prior research up to the present day, though it does have limitations such as being unable to backtrack.

As a result of the variance possible in the agent’s decision making we run the agent on each level five times and then calculate the average results across these five runs for all of the agent extracted features.

\subsection{Heatmap Generation and Distribution Analysis}

In order to calculate the fitness independence selection criteria, as well as some of the results visualisations, the expressive range plot needs to be divided into a grid so that the average fitness of levels in each grid location can be calculated. Our approach for this is to divide each metric pair ERA plot into a 20 by 20 grid, with the range of values for each cell determined by subtracting the minimum value found for a metric from the maximum value and then dividing by 20. This approach ensures that all levels can be placed within the plot, and that no metric specific ranges have to be chosen. However, it also means that the distribution within the plot is heavily influenced by the presence of outlying metric values, something we discuss further in Section \ref{sect_limitations}. 

\subsection{Data and Visualisations Presented}

In the following section we present several results tables and visualisations to illustrate the important and interesting findings from these experiments. First, we present the ERA visualisations of the highest and lowest ranked metric pairings in terms of their average rank across the three selection criteria. These are presented both with the playability heatmaps to show how playable levels are distributed in the space, along with ERA visualisations of the full level corpus with different generators highlighted in different colours. These are presented to highlight some of the features of strong and weak metric pairings when using our selection criteria.

Second, we present top 5 ranked metric pairings for each selection criteria in terms of their values, as well as the top 5 ranked pairings in terms of their average rank across the three other ranks. These are primarily presented to demonstrate what the final data obtained from applying the selection criteria looks like, and to highlight patterns in the ranks obtained by high performing metric pairs across the three criteria. This set of 20 pairings is also used as our set of high performing metric pairs, to be used in the final two figures which illustrate patterns found in metric pairings which obtained high rankings in at least one criteria.

Finally, we present both a histogram of the appearance rate of each individual metric in the list of top high performing pairings, as well as a table enumerating the count of pairings in this list which were either both agent selected or structural metrics, or were composed of one of each. These are presented to highlight valuable trends in the makeup of high performing pairs, both to present avenues for future work but also to highlight weaknesses in the presented implementation.

\subsection{Hardware \& Software Used}

In order to extract and save the metrics for each of the Mario levels included in the experiments, custom Java code was used to interface with the Mario AI Benchmark. Code available at \url{github.com/KrellFace/mario\_metric\_extraction}. The calculation of the selection criteria themselves as well as the generation of visualisations was done in Python, with code available at \url{github.com/KrellFace/ERA\_Metric\_Evaluation}.

All software was run on a Dell laptop with an i5-10310U CPU with 16.0 GB of RAM. The process of metric extraction from the Mario AI Benchmark took approximately 12 hours, and the calculation of the selection criteria took approximately 1 hour and 30 minutes.

\section{Results and Discussion}\label{sect_results}
\begin{table*}
  \caption{Top 5 metric pairs for each selection criteria, as well as for mean average rank across all three criteria. All values rounded to 3sf}
  \label{tab_mcrank}
\begin{tabular}{cllllllll}
\cline{3-9}
\multicolumn{1}{l}{}                                                                                              & \multicolumn{1}{l|}{\textbf{}}                      & \multicolumn{2}{l|}{\textbf{FI}}                                                                    & \multicolumn{2}{l|}{\textbf{MC}}                                                                    & \multicolumn{2}{l|}{\textbf{AMC}}                                                                   & \multicolumn{1}{l|}{\textbf{}}                    \\ \hline
\multicolumn{1}{|c|}{}                                                                                            & \multicolumn{1}{l|}{\textbf{Metric Pair}}           & \multicolumn{1}{l|}{\textbf{Score}}                & \multicolumn{1}{l|}{\textbf{Rank}}             & \multicolumn{1}{l|}{\textbf{Score}}                & \multicolumn{1}{l|}{\textbf{Rank}}             & \multicolumn{1}{l|}{\textbf{Score}}                & \multicolumn{1}{l|}{\textbf{Rank}}             & \multicolumn{1}{l|}{\textbf{Average Rank}}        \\ \hline
\multicolumn{1}{|c|}{}                                                                                            & \multicolumn{1}{l|}{Density-JumpCount}              & \multicolumn{1}{l|}{\cellcolor[HTML]{FFFFC7}0.517} & \multicolumn{1}{l|}{\cellcolor[HTML]{FFFFC7}1} & \multicolumn{1}{l|}{0.885}                         & \multicolumn{1}{l|}{30}                        & \multicolumn{1}{l|}{0.518}                         & \multicolumn{1}{l|}{77}                        & \multicolumn{1}{l|}{36}                           \\ \cline{2-9} 
\multicolumn{1}{|c|}{}                                                                                            & \multicolumn{1}{l|}{EnemyCount-JumpCount}           & \multicolumn{1}{l|}{\cellcolor[HTML]{FFFFC7}0.514} & \multicolumn{1}{l|}{\cellcolor[HTML]{FFFFC7}2} & \multicolumn{1}{l|}{0.494}                         & \multicolumn{1}{l|}{101}                       & \multicolumn{1}{l|}{0.521}                         & \multicolumn{1}{l|}{74}                        & \multicolumn{1}{l|}{59}                           \\ \cline{2-9} 
\multicolumn{1}{|c|}{}                                                                                            & \multicolumn{1}{l|}{Density-AverageY}               & \multicolumn{1}{l|}{\cellcolor[HTML]{FFFFC7}0.496} & \multicolumn{1}{l|}{\cellcolor[HTML]{FFFFC7}3} & \multicolumn{1}{l|}{0.985}                         & \multicolumn{1}{l|}{3}                         & \multicolumn{1}{l|}{0.551}                         & \multicolumn{1}{l|}{39}                        & \multicolumn{1}{l|}{15}                           \\ \cline{2-9} 
\multicolumn{1}{|c|}{}                                                                                            & \multicolumn{1}{l|}{JumpCount-TotalEnemyDeaths}     & \multicolumn{1}{l|}{\cellcolor[HTML]{FFFFC7}0.479} & \multicolumn{1}{l|}{\cellcolor[HTML]{FFFFC7}4} & \multicolumn{1}{l|}{0.345}                         & \multicolumn{1}{l|}{130}                       & \multicolumn{1}{l|}{0.541}                         & \multicolumn{1}{l|}{47}                        & \multicolumn{1}{l|}{60.3}                         \\ \cline{2-9} 
\multicolumn{1}{|c|}{\multirow{-5}{*}{\textbf{\begin{tabular}[c]{@{}c@{}}Top 5\\ FI\\ Rank\end{tabular}}}}        & \multicolumn{1}{l|}{PipeCount-TotalEnemyDeaths}     & \multicolumn{1}{l|}{\cellcolor[HTML]{FFFFC7}0.462} & \multicolumn{1}{l|}{\cellcolor[HTML]{FFFFC7}5} & \multicolumn{1}{l|}{0.406}                         & \multicolumn{1}{l|}{122}                       & \multicolumn{1}{l|}{0.5831}                        & \multicolumn{1}{l|}{15}                        & \multicolumn{1}{l|}{47.3}                         \\ \hline
\multicolumn{1}{l}{}                                                                                              &                                                     &                                                    &                                                &                                                    &                                                &                                                    &                                                &                                                   \\ \hline
\multicolumn{1}{|c|}{}                                                                                            & \multicolumn{1}{l|}{EmptySpaceCount-MaxJumpAirTime} & \multicolumn{1}{l|}{0.0564}                        & \multicolumn{1}{l|}{153}                       & \multicolumn{1}{l|}{\cellcolor[HTML]{FFFFC7}0.997} & \multicolumn{1}{l|}{\cellcolor[HTML]{FFFFC7}1} & \multicolumn{1}{l|}{0.583}                         & \multicolumn{1}{l|}{17}                        & \multicolumn{1}{l|}{57}                           \\ \cline{2-9} 
\multicolumn{1}{|c|}{}                                                                                            & \multicolumn{1}{l|}{Speed-MaxJumpAirTime}           & \multicolumn{1}{l|}{0.124}                         & \multicolumn{1}{l|}{132}                       & \multicolumn{1}{l|}{\cellcolor[HTML]{FFFFC7}0.985} & \multicolumn{1}{l|}{\cellcolor[HTML]{FFFFC7}2} & \multicolumn{1}{l|}{0.4521}                        & \multicolumn{1}{l|}{128}                       & \multicolumn{1}{l|}{87.3}                         \\ \cline{2-9} 
\multicolumn{1}{|c|}{}                                                                                            & \multicolumn{1}{l|}{Density-AverageY}               & \multicolumn{1}{l|}{0.496}                         & \multicolumn{1}{l|}{3}                         & \multicolumn{1}{l|}{\cellcolor[HTML]{FFFFC7}0.985} & \multicolumn{1}{l|}{\cellcolor[HTML]{FFFFC7}3} & \multicolumn{1}{l|}{0.551}                         & \multicolumn{1}{l|}{39}                        & \multicolumn{1}{l|}{15}                           \\ \cline{2-9} 
\multicolumn{1}{|c|}{}                                                                                            & \multicolumn{1}{l|}{Density-ClearColumns}           & \multicolumn{1}{l|}{0.296}                         & \multicolumn{1}{l|}{67}                        & \multicolumn{1}{l|}{\cellcolor[HTML]{FFFFC7}0.983} & \multicolumn{1}{l|}{\cellcolor[HTML]{FFFFC7}4} & \multicolumn{1}{l|}{0.286}                         & \multicolumn{1}{l|}{148}                       & \multicolumn{1}{l|}{73}                           \\ \cline{2-9} 
\multicolumn{1}{|c|}{\multirow{-5}{*}{\textbf{\begin{tabular}[c]{@{}c@{}}Top 5\\ MC\\ Rank\end{tabular}}}}        & \multicolumn{1}{l|}{KillsByStomp-AverageY}          & \multicolumn{1}{l|}{0.321}                         & \multicolumn{1}{l|}{53}                        & \multicolumn{1}{l|}{\cellcolor[HTML]{FFFFC7}0.983} & \multicolumn{1}{l|}{\cellcolor[HTML]{FFFFC7}5} & \multicolumn{1}{l|}{0.536}                         & \multicolumn{1}{l|}{54}                        & \multicolumn{1}{l|}{37.3}                         \\ \hline
\multicolumn{1}{l}{}                                                                                              &                                                     &                                                    &                                                &                                                    &                                                &                                                    &                                                &                                                   \\ \hline
\multicolumn{1}{|c|}{}                                                                                            & \multicolumn{1}{l|}{EmptySpaceCount-JumpCount}      & \multicolumn{1}{l|}{0.122}                         & \multicolumn{1}{l|}{133}                       & \multicolumn{1}{l|}{0.465}                         & \multicolumn{1}{l|}{106}                       & \multicolumn{1}{l|}{\cellcolor[HTML]{FFFFC7}0.610} & \multicolumn{1}{l|}{\cellcolor[HTML]{FFFFC7}1} & \multicolumn{1}{l|}{80}                           \\ \cline{2-9} 
\multicolumn{1}{|c|}{}                                                                                            & \multicolumn{1}{l|}{BlockCount-TotalEnemyDeaths}    & \multicolumn{1}{l|}{0.336}                         & \multicolumn{1}{l|}{44}                        & \multicolumn{1}{l|}{0.621}                         & \multicolumn{1}{l|}{83}                        & \multicolumn{1}{l|}{\cellcolor[HTML]{FFFFC7}0.609} & \multicolumn{1}{l|}{\cellcolor[HTML]{FFFFC7}2} & \multicolumn{1}{l|}{43}                           \\ \cline{2-9} 
\multicolumn{1}{|c|}{}                                                                                            & \multicolumn{1}{l|}{BlockCount-AverageY}            & \multicolumn{1}{l|}{0.321}                         & \multicolumn{1}{l|}{52}                        & \multicolumn{1}{l|}{0.372}                         & \multicolumn{1}{l|}{125}                       & \multicolumn{1}{l|}{\cellcolor[HTML]{FFFFC7}0.606} & \multicolumn{1}{l|}{\cellcolor[HTML]{FFFFC7}3} & \multicolumn{1}{l|}{60}                           \\ \cline{2-9} 
\multicolumn{1}{|c|}{}                                                                                            & \multicolumn{1}{l|}{Contiguity-TotalEnemyDeaths}    & \multicolumn{1}{l|}{0.334}                         & \multicolumn{1}{l|}{45}                        & \multicolumn{1}{l|}{0.737}                         & \multicolumn{1}{l|}{54}                        & \multicolumn{1}{l|}{\cellcolor[HTML]{FFFFC7}0.606} & \multicolumn{1}{l|}{\cellcolor[HTML]{FFFFC7}4} & \multicolumn{1}{l|}{34.3}                         \\ \cline{2-9} 
\multicolumn{1}{|c|}{\multirow{-5}{*}{\textbf{\begin{tabular}[c]{@{}c@{}}Top 5\\ AMC \\ Rank\end{tabular}}}}      & \multicolumn{1}{l|}{BlockCount-JumpCount}           & \multicolumn{1}{l|}{0.359}                         & \multicolumn{1}{l|}{30}                        & \multicolumn{1}{l|}{0.522}                         & \multicolumn{1}{l|}{99}                        & \multicolumn{1}{l|}{\cellcolor[HTML]{FFFFC7}0.604} & \multicolumn{1}{l|}{\cellcolor[HTML]{FFFFC7}5} & \multicolumn{1}{l|}{44.7}                         \\ \hline
\multicolumn{1}{l}{}                                                                                              &                                                     &                                                    &                                                &                                                    &                                                &                                                    &                                                &                                                   \\ \hline
\multicolumn{1}{|c|}{}                                                                                            & \multicolumn{1}{l|}{Density-AverageY}               & \multicolumn{1}{l|}{0.496}                         & \multicolumn{1}{l|}{3}                         & \multicolumn{1}{l|}{0.985}                         & \multicolumn{1}{l|}{3}                         & \multicolumn{1}{l|}{0.551}                         & \multicolumn{1}{l|}{39}                        & \multicolumn{1}{l|}{\cellcolor[HTML]{FFFFC7}15}   \\ \cline{2-9} 
\multicolumn{1}{|c|}{}                                                                                            & \multicolumn{1}{l|}{PipeCount-Density}              & \multicolumn{1}{l|}{0.365}                         & \multicolumn{1}{l|}{26}                        & \multicolumn{1}{l|}{0.947}                         & \multicolumn{1}{l|}{20}                        & \multicolumn{1}{l|}{0.574}                         & \multicolumn{1}{l|}{22}                        & \multicolumn{1}{l|}{\cellcolor[HTML]{FFFFC7}22.7} \\ \cline{2-9} 
\multicolumn{1}{|c|}{}                                                                                            & \multicolumn{1}{l|}{Density-JumpEntropy}            & \multicolumn{1}{l|}{0.456}                         & \multicolumn{1}{l|}{6}                         & \multicolumn{1}{l|}{0.896}                         & \multicolumn{1}{l|}{29}                        & \multicolumn{1}{l|}{0.530}                         & \multicolumn{1}{l|}{60}                        & \multicolumn{1}{l|}{\cellcolor[HTML]{FFFFC7}31.7} \\ \cline{2-9} 
\multicolumn{1}{|c|}{}                                                                                            & \multicolumn{1}{l|}{ClearColumns-AverageY}          & \multicolumn{1}{l|}{0.396}                         & \multicolumn{1}{l|}{17}                        & \multicolumn{1}{l|}{0.966}                         & \multicolumn{1}{l|}{12}                        & \multicolumn{1}{l|}{0.522}                         & \multicolumn{1}{l|}{71}                        & \multicolumn{1}{l|}{\cellcolor[HTML]{FFFFC7}33.3} \\ \cline{2-9} 
\multicolumn{1}{|c|}{\multirow{-5}{*}{\textbf{\begin{tabular}[c]{@{}c@{}}Top 5 \\ Average \\ Rank\end{tabular}}}} & \multicolumn{1}{l|}{Contiguity-TotalEnemyDeaths}    & \multicolumn{1}{l|}{0.334}                         & \multicolumn{1}{l|}{45}                        & \multicolumn{1}{l|}{0.737}                         & \multicolumn{1}{l|}{54}                        & \multicolumn{1}{l|}{0.606}                         & \multicolumn{1}{l|}{4}                         & \multicolumn{1}{l|}{\cellcolor[HTML]{FFFFC7}34.3} \\ \hline
\end{tabular}
\end{table*}

\begin{figure*}
  \centering
\includegraphics[width=\linewidth]{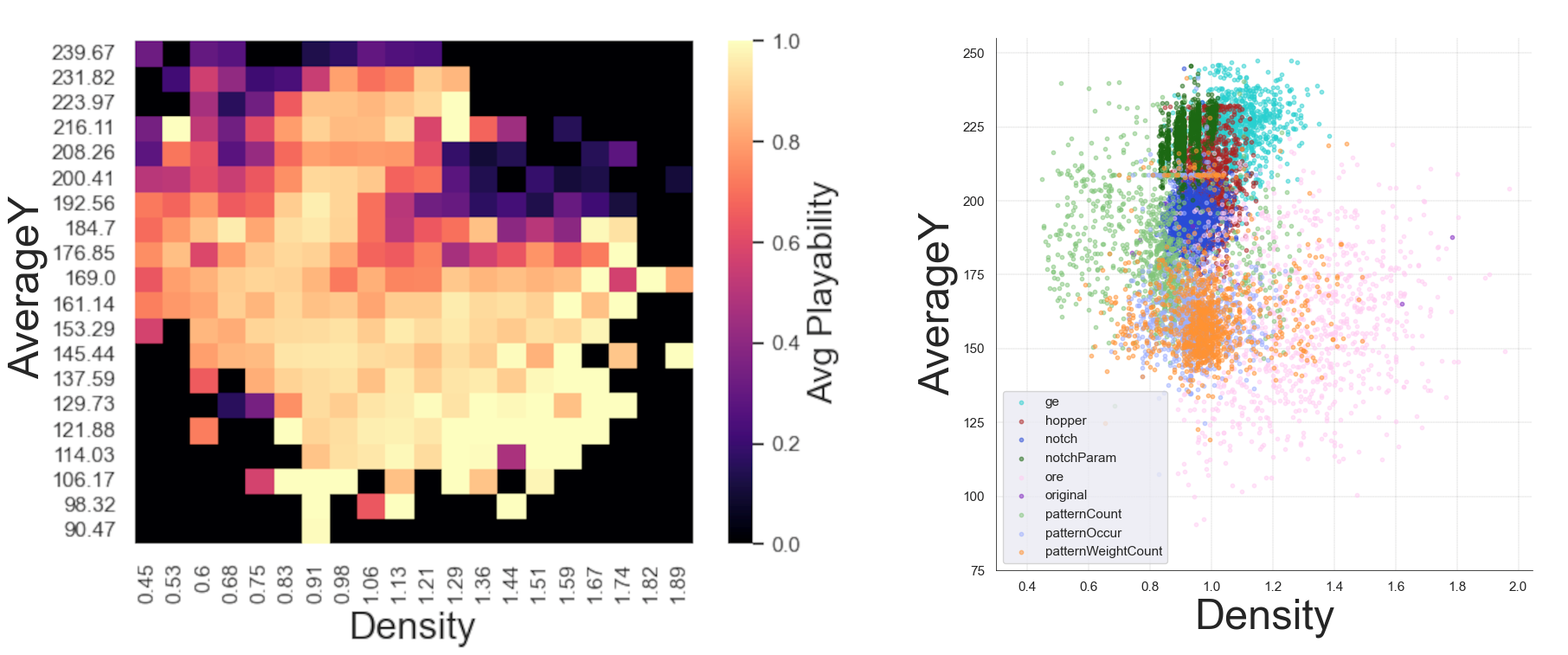}
  \caption{Heatmap of highest ranked metric pair, Density \& AverageY,  in terms of average rank. Presented with both the fitness heatmap, and the derived ERA plot when using the metrics to project the level corpus, with each generator highlighted in a different colour}
  \label{fig_best}
\end{figure*}

\begin{figure*}
  \centering
\includegraphics[width=\linewidth]{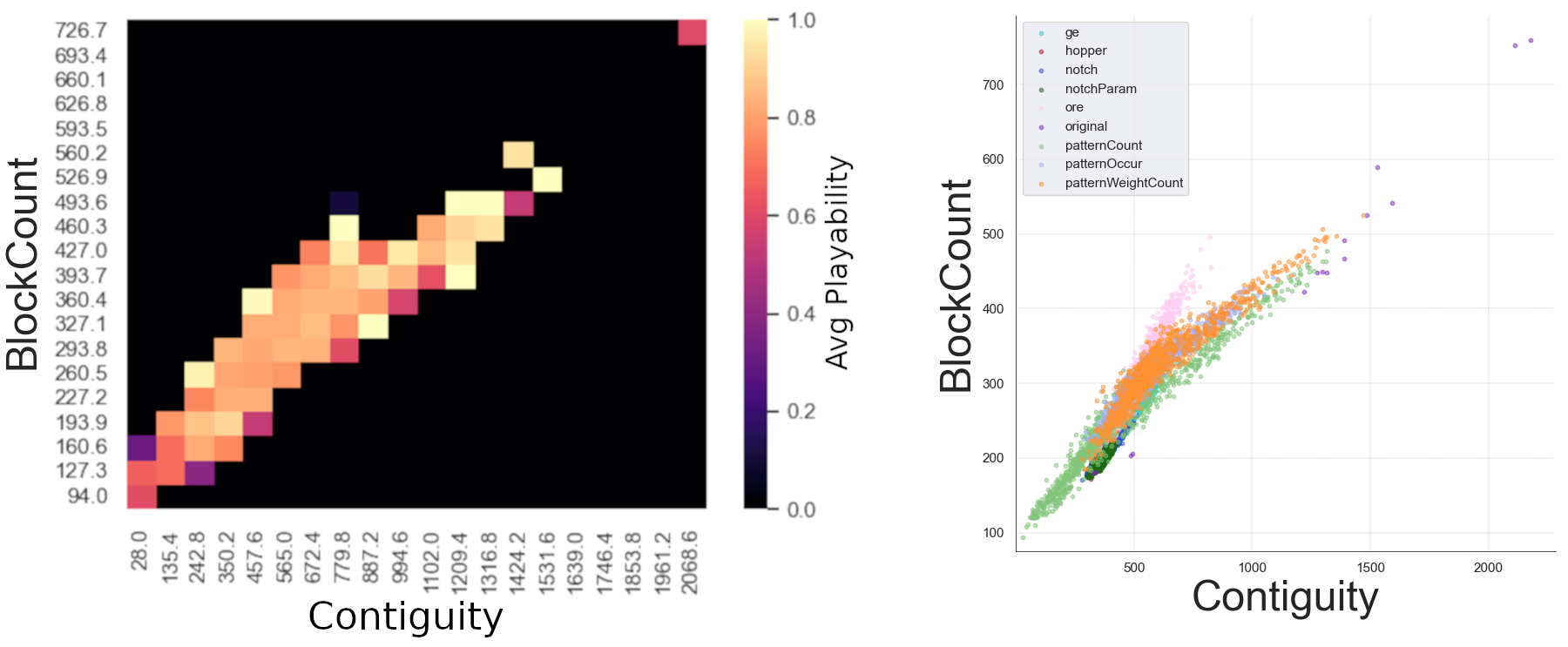}
  \caption{Heatmap of lowest ranked metric pair, Contiguity \& BlockCount,  in terms of average rank. Presented with both the fitness heatmap, and the derived ERA plot when using the metrics to project the level corpus, with each generator highlighted in a different colour}
  \label{fig_worst}
\end{figure*}

\begin{table}
  \caption{Frequency of alternative combinations of structural and agent extracted features across the 5 top ranked pairs for each criteria and for average rank}
  \label{tab_metricfreq}
\begin{tabular}{l|lllll|}
\cline{2-6}
                                             & \multicolumn{5}{c|}{\textbf{Count}}                                                                                                                                   \\ \hline
\multicolumn{1}{|l|}{\textbf{Pair Category}} & \multicolumn{1}{l|}{\textbf{Avg Rank}} & \multicolumn{1}{l|}{\textbf{FI}} & \multicolumn{1}{l|}{\textbf{MC}} & \multicolumn{1}{l|}{\textbf{AMC}} & \textbf{Total} \\ \hline
\multicolumn{1}{|l|}{Structural-Structural}  & \multicolumn{1}{l|}{1}                     & \multicolumn{1}{l|}{0}           & \multicolumn{1}{l|}{1}           & \multicolumn{1}{l|}{0}            & \textbf{2}     \\ \hline
\multicolumn{1}{|l|}{Structural-Agent}       & \multicolumn{1}{l|}{4}                     & \multicolumn{1}{l|}{4}           & \multicolumn{1}{l|}{2}           & \multicolumn{1}{l|}{5}            & \textbf{15}    \\ \hline
\multicolumn{1}{|l|}{Agent-Agent}            & \multicolumn{1}{l|}{0}                     & \multicolumn{1}{l|}{1}           & \multicolumn{1}{l|}{2}           & \multicolumn{1}{l|}{0}            & \textbf{3}     \\ \hline
\end{tabular}
\end{table}

\begin{figure}
  \centering
  \includegraphics[width=\linewidth]{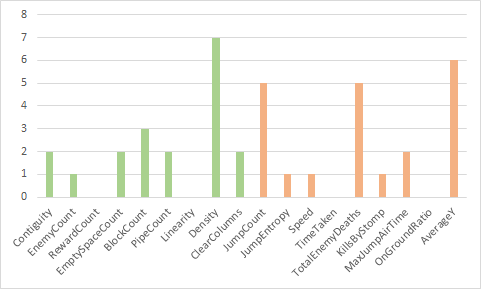}
  \caption{Frequency of appearance of each metric in the top 5 pairs for each selection criteria and average rank. Structural metrics in green, agent-extracted in orange}
  \label{fig_relfreq}
\end{figure}

In this section we discuss the approach we have presented, as well as the insights into it that were gained from our pilot experiments. Overall we find many reasons to be enthusiastic about the approach we have presented. The process for calculating the section criteria for metric pairs is straightforward and relatively fast, and the insights that can be gained from using it can be both useful and also unobvious. We are confident that using them would likely lead to more informative and fair ERA plots than making an uninformed choice. To get an idea of this we can look at the visualisations for the best and worst performing metric pairs in terms of their average ranks in Figures \ref{fig_best} and \ref{fig_worst}. Not only is it easier to distinguish between the generative spaces of the different generators in the best performing metric pair in Figure \ref{fig_best}, but as a result of the FI and AMC selection criteria one can be more confident that what appears to be diversity is both useful in terms of its fitness, and also represents diversity across multiple alternative metrics.

One of the interesting patterns in our pilot results is the relationship, or lack of one, between the performance of metric pairs across the three criteria. As we can see in Table \ref{tab_mcrank}, metric pairs that ranked highly in one selection criteria often ranked very poorly for other selection criteria. We expected the FI and MC criteria to be correlated to an extent as they both prioritise the wide distribution of levels in the space. We also expected MC and AMC criteria to be correlated as the more uncorrelated the two metrics are the more likely it is that one of them will have a strong correlation with a given alternative metric. These correlation effects do not appear strong enough in this domain to ensure the existence of metric pairs which are high performing in all criteria. For an example of this, the 2nd ranked metric pair in terms of average rank, PipeCount-Density, does not appear higher than rank 20 for any individual criteria. It is possible however, that alternative untested metrics or improved designs of tested ones could attain good performance on all three criteria.

As we can see in Figure \ref{fig_relfreq}, the individual metrics we investigated were widely distributed in terms of how often they appeared in high ranked metric pairs. We looked at the Top 5 ranked pairs for each criteria and the Top 5 averaged ranked, giving a total of 20 high performing pairs and 40 constituent metrics. At the high end was the density metric which appeared in 7 of 20 possible pairs. At the low end, four metrics did not appear at all. While this sample is too small and domain specific to make any general statements, the level of skew is interesting and implies that we as researchers should be more cautious about making an unconsidered choice about which metrics to use. 

A thorough investigation of what the underlying causes of individual metrics underperforming is outside the scope of this work. However some causes were immediately apparent, such as the damaging effect of outliers on how we calculated the FI selection criteria. Extreme outliers cause the non outlying levels to be concentrated in the heatmap, as the heat maps dimensions are determined by the total range of the metric values, which harms their level of distribution and their subsequent score. This was most obvious in the MaxJumpAirTime metric, in which one level Mario was able to achieve a value of 148 frames, well outside of the rest of the values for the remaining hundreds of levels whose values were found in the range of 0 to 62. This high value was achieved due to the unique design of the level which was so highly populated by enemies that the most efficient route for the A* agent was to bounce on them for the majority of the level, leading to long periods of time not spent on the ground. This directly caused the low ranking of any metric pair using it in terms of the FI score, as we can see in Table\ref{tab_mcrank} with the otherwise strongly performing ‘EmptySpaceCount-MaxJumpAirTime’ metric pair which is top ranked for the MC selection criteria. While this is interesting, it should also caution us against using this method of calculating Fitness Independence again, without further considering how to handle outliers.

An interesting finding was the prevalence of metric pairs involving a combination of structural and agent evaluated features in high ranking pairs. As we see in Table \ref{tab_metricfreq}, 15 out of the 20 high scoring pairs conformed to this format. Statistically there is a bias towards structural-agent pairs, as due to us not testing pairs of the same metric a slight majority of 81 of the 153 total pairs were structural-agent pairs. However the skew was more extreme than can be explained by this bias, and it also makes intuitive sense, as the way a game playing agent interacts with the level as a series of required actions is very different to structurally analysing it as a matrix of tiles. We expect this to mean they are less likely to be mutually correlated, and therefore more likely to score well on the AMC criteria. This result suggests that ERA implementations could benefit from calculating each metric in their pairs using differing approaches to increase the likelihood that the metrics are performant in terms of our selection criteria. Though we should be careful to not extrapolate too far from this single data point, as it could easily be a quirk of the specific metrics we selected to evaluate.

\section{Limitations and Future Work}\label{sect_limitations}

While we are excited about the concept and high level approach we have presented, it has several weaknesses and limitations that could be usefully improved upon. Additionally there is much further work that could be done to both expand on the ideas we have presented, and to generally advance the research domain of understanding and visualising PCG system generative spaces.

An important avenue of future work is to explore alternative ways of calculating the selection criteria that were introduced in this paper. Rank correlation calculated using Spearman’s Rho for the MC and AMC criteria is fast and efficient, but it has the limitation of only detecting monotonic relationships between metrics, and could not be able to detect more complex non-linear relationships. Future work could usefully explore more complex but computationally intensive alternatives such as the use of smoothing splines. There are also alternatives for calculating the FI metric. A simple alternative could be to calculate the maximum level fitness found in a grid cell rather than the average which could be fairer and more valid, as what we are interested in with this criteria is that it is possible to find levels widely within the ERA plot, not whether a high proportion of them have already been found. 

This area could also benefit highly from further exploration and introspection on what additional selection criteria would be valuable for metric pairs. One which could be valuable but was outside of scope for this paper is that of emergence, meaning that values for metrics are not determined by how the generator is parameterised. This was argued as important by Smith et al in the paper that first introduced ERA and we think it would still be valuable for making ERA more informative, as well as making comparisons between generative spaces fairer. Operationalising this criteria would be complex as one would need sets of levels generated from multiple parameterisations of the same generator, but future work could explore an approach similar to RERA, developed by Cook et al for the Danesh platform \cite{cook2021}.

A conceptual weakness of both our approach and ERA more generally, is the unknowability of the underlying possibility space. As we discussed in Section \ref{sect_pspace} the process of tuning a PCG system can be thought of as trying to optimise the size and location of the system’s generative space within the wider possibility space of possible content. When ERA is being used to quantify the search space exploration of a generator, what researchers want to know is how much of the underlying possibility space is being explored. However, there has been little prior work which aims to understand or quantify the possibility spaces which encompass generative spaces. Future work to explore methods for getting a PCG system agnostic view of underlying possibility spaces could be highly beneficial for both robust ERA, but also PCG research in general. Quality-Diversity based PCG approaches \cite{gravina2019} provide a possible avenue for this as they are well optimised for maximising exploration. As they also create projections of the search space using metrics in an identical way to ERA there would need to be careful consideration of what metrics to choose. This work’s selection criteria could potentially be deployed for selecting metrics to increase the possibility space exploration of QD search, something we intend to investigate in future.

A practical weakness of the approach we have presented is simply the time it takes to identify the best metric pairs. While the computation time was relatively short, it obviously takes far longer than conducting the ERA itself. This makes it substantially less appealing for practical ERA applications, and would only get less appealing as the number of candidate metrics increased, though increasing the number of candidates would also arguably increase the value of our selection criteria by helping a designer to filter them. In future work we would like to explore whether metrics can be found which are highly performant across multiple game domains of similar genre. This could synergise well with papers discussed previously which explored the relationship between metrics and player perceptions \cite{marino2021,summerville2017,herve2021}. Metric pairs that scored highly in multiple domains using our criteria, while also correlating with human perceptions of quality or diversity would have the potential to be robustly useful in novel PCG domains without the need for evaluation.

\section{Conclusion}\label{sect_conclusion}

In this paper we have presented our approach for enhancing implementations of ERA through the use of metric selection criteria to enable designers to make a more informed choice on which pair of metrics is most appropriate for their domain. We also presented our implementation of these criteria and the results and insights we obtained from applying them to a pre-existing corpus of generated levels. We find that using these criteria can give valuable insights on how best to produce ERA visualisations of a domain, as well as how best to design metrics for conducting ERA. We hope that this work might form the basis of future work exploring novel methods for selecting the best pairs of metrics for expressive visualisation within PCG. 

\bibliographystyle{ACM-Reference-Format}
\bibliography{cameraReadyBib}

\end{document}